# Testbed Development: An Intelligent O-RAN based Cell-Free MIMO Network

Yi Chu, Mostafa Rahmani, Josh Shackleton, David Grace,
Kanapathippillai Cumanan, Hamed Ahmadi, Alister Burr
School of Physics, Engineering and Technology, University of York, UK

*Abstract*—Cell-free multiple input multiple output (CF-MIMO) systems improve spectral and energy efficiencies using distributed access points (APs) to provide reliable service across an area equivalent to multiple conventional cells. This paper presents a novel design and implementation of a CF-MIMO network leveraging the open radio access network (O-RAN) architecture based testbed to enhance the performance of interference-prone user. The proposed prototype is developed based on open source software components and unlike many other prototypes, our testbed is able to serve commercial 5G user equipment (UE). The RAN intelligent controller (RIC) allows the cell-free (CF) network to access the embedded artificial intelligence and benefit from the network optimisation techniques that O-RAN brings. The testbed includes an intelligent antenna association xApp which determines the antenna group that serves each UE based on the live key performance measurements. The paper demonstrates the deployment and operation of the CF network and the xApp and discusses how the CF networks can benefit from the O-RAN architecture.

## I. INTRODUCTION

The rapid evolution of wireless communication technologies necessitates innovative approaches to meet the increasing demands for throughput, coverage, and user experiences. Traditional cellular networks often struggle to provide adequate service for cell-edge users, or interference-prone users, primarily due to attenuation and interference from adjacent base stations (BSs). To address these challenges, the concept of cell-free multiple input multiple output (CF-MIMO) networks has emerged as a promising solution [1].

CF-MIMO technology offers a paradigm shift from traditional cellular networks by eliminating cell boundaries and enabling a cooperative network of access points (APs) that jointly serve all user equipment (UE). This approach provides macro-diversity, enhancing coverage and throughput, particularly for interference-prone users. In a CF-MIMO network, a large number of distributed APs, each potentially equipped with multiple antennas, are connected to a central processing unit (CPU) via a fronthaul network, enabling coherent transmission and reception. The CPU operates the system in a distributed MIMO fashion, significantly improving the signal-to-noise ratio (SNR) and signal-interference-plus-noise ratio (SINR) for users, regardless of their locations. Recent advancements in the implementation of CF-MIMO systems have demonstrated their potential in both centralized and distributed architectures. Centralized architectures, leveraging advanced techniques like minimum mean-square error (MMSE) or zero-forcing (ZF) combining, maximize spectral efficiency and reduce fronthaul signalling compared to standard distributed approaches, while distributed architectures offer flexibility for ultra-dense networks depending on access point cooperation [2].

The evolution towards 6G requires significant improvements such as network capacity, spectral efficiency, energy consumption and reliability [3]. CF-MIMO has been considered as a promising architecture for 6G given its unique benefits to interference-prone users [4]. Open radio access network (O-RAN) as another enabling technology for future mobile networks provides open interfaces which allow multi-vendor network deployment [5]. O-RAN also disaggregates the RAN into radio unit (RU), distributed unit (DU) and central units (CU). When multiple RUs are connected to the same DU, the DU naturally becomes the CPU of the CF architecture which allows the UE to be served by RUs from different vendors simultaneously. The RIC [6] and the customisable RICapps it hosts enable artificial intelligence (AI)-based optimisation including energy efficiency, traffic steering and dynamic slicing. The network can also benefit from optimisation methods unique to the CF architecture such as serving antenna group association, pilot assignment, UE power control, UE precoding/combiner techniques, backhaul/fronthaul optimization, and DU/CPU location optimisation.

### A. Available Cell-Free Network Testbeds

The mutual benefits and complementary features of the CF and O-RAN architectures have motivated us to develop a prototype to evaluate this hybrid approach. Based on the information available, only the O-RAN compatible fronhaul has been introduced by some of the available CF testbeds. The RIC and E2 interface which are the main enabler of AI, have not been implemented yet. This section reviews key works that have contributed to the development and understanding of CF-MIMO systems.

The work in [7] presents the implementation of CF-MIMO combined with dynamic time division duplex (TDD) using the OpenAirInterface (OAI) 5G stack. Dynamic TDD which allows cells to schedule uplink (UL) and downlink (DL) slots based on traffic demands, has been integrated with CF-MIMO to mitigate cross-link interference. This testbed has all radios (RU and UE) synchronised to the same reference source and demonstrated the 5G physical layer (PHY) with a fixed modulation and coding scheme (MCS). The authors in [8] demonstrate a distributed CF-MIMO testbed with video streaming. This testbed uses distributed computing without a CPU, thereby reducing fronthaul capacity demands. The testbed consists of 10 APs and 4 UEs (all implemented with software defined radios (SDRs)), each with a single antenna. The testbed has demonstrated multi-user (MU)-MIMO with UEs served by multiple RUs on the same resource, with the Labview 5G PHY stack. The work has been further extended to facilitate the 5G/6G-based private networks suitable for industrial internet of things applications [9]. The work proposes a novel orthogonal frequency division multiplexing (OFDM) precoding algorithm, which allows APs in a user-centric cluster to maximize capacity while minimizing interference to neighbouring UEs.

TABLE I: Comparison of CF-MIMO System Implementations

| Reference | O-RAN Architecture | UE type | Synchronization | MAC Scheduler | Protocol Stack | Modulation/Coding Scheme |
|---|---|---|---|---|---|---|
| [7] | No | SDR | RUs and UEs are synchronised with an Octoclock | Unspecified | OAI 5G PHY stack | Fixed MCS (MCS=10, 16QAM), No MU-MIMO |
| [8] | No | SDR | RUs are synchronised with PTP | Continuous transmission | Only PHY implemented in Labview | Fixed with QPSK, 16QAM and 64QAM |
| [9] | No | SDR | RUs are synchronised with PTP | Unspecified | 4G stack implemented in Labview | 16QAM maximum |
| [10] | No | similar radio as RU | UEs and RUs are synchronised with PTP and GPS | Unspecified | 5G PHY stack | Adaptive MCS with non-standard 48-port DMRS |
| [11] | Fronthaul only | similar radio as RU | UEs and RUs are synchronised with PTP and GPS | Unspecified | Unspecified | Unspecified |
| [12] | No | SDR | RUs are synchronised with PTP | Unspecified | 5G PHY stack | 16QAM maximum |
| Ours | RIC and E2 included | Commercial 5G UE | RUs are synchronised with an Octoclock | Defined for coexistence of SU and MU-MIMO | Complete 5G stack | Adaptive MCS |

The work in [10] proposes a full-spectrum CF-RAN. Key innovations include a scalable CF-MIMO architecture and a CF-MIMO testbed with 64 distributed antennas, using commercial radios. The testbed has all radios synchronised with precision time protocol (PTP) and demonstrated 5G MU-MIMO PHY with adaptive MCS and customised 48-port demodulation reference signal (DMRS) signals. The authors in [11] investigate the downlink coherent multi-user transmission in CF-MIMO systems, focusing on over-the-air (OTA) reciprocity calibration and phase synchronization. A testbed including 4 remote radio units (RRUs) and 4 UEs has been developed to evaluate such approach. Similar to the work in [10], all radios (RRUs and UEs) are synchronised with PTP. In [12], the authors propose a cloud-based CF distributed massive MIMO system. The system addresses challenges in synchronization, calibration, and real-time processing, using a 128 x 128 MIMO setup. The testbed demonstrates 5G PHY with high spectral efficiency and throughput.

*B. Contributions*

With the review of implemented CF-MIMO prototypes, it is clear that the focus has been on novel low technology readiness level (TRL) PHY technologies. Table I offers a concise summary of these works, emphasizing the areas they address and the critical gaps to be investigated. With the latest development of the open source O-RAN software, we have identified the opportunity to build a mid-TRL testbed with the hybrid CF and O-RAN architecture to demonstrate the benefits of combining the two 6G enabling technologies. Our work aims at developing an intelligent private 5G network to serve commercial-off-the-shelf (COTS) UEs while addressing the gaps identified in Table I. Our contributions are threefold:

- **Full O-RAN compatible 5G software stack:** unlike other CF-MIMO prototypes which mostly only include PHY, our testbed includes all software components required to serve COTS UEs, such as a full 5G RAN stack, a standalone (SA) 5G core, and a RIC with onboarded xApps. The RAN is designed to be scalable with the number of RUs with the support of MU-MIMO for the UL. To the best of our knowledge, we have not yet noticed any CF-MIMO prototypes which can serve COTS 5G UEs with MU-MIMO. Our testbed is also the first implementation of CF-MIMO on the O-RAN architecture.
- **Intelligence embedded at the RIC:** designed based on the O-RAN architecture, our testbed naturally inherits the AI driven network optimisation from the RIC, which is absent from any existing CF-MIMO prototypes. Many existing O-RAN optimisation methods can be directly applied such as switching RUs on/off for energy saving and proactive UE handover for traffic steering. Novel optimisations for the CF architecture can also be developed, such as UE serving antenna group association (which will be demonstrated later), UE power control and DU/CPU location optimisation.
- **Medium access control (MAC) scheduler:** the MAC scheduler is another critical gap across the existing CF-MIMO prototypes which is essential for serving COTS UEs. The UEs send scheduling requests (SRs) with different patterns and the scheduler needs to allocate appropriate resources from the resource grid, rather than continuous transmission on the same set of physical resource blocks (PRBs), which is commonly used across existing prototypes. The scheduler of our testbed allows coexistence of single-user (SU)/MU-MIMO and provides the PHY the corresponding information for decoding the UE signal.

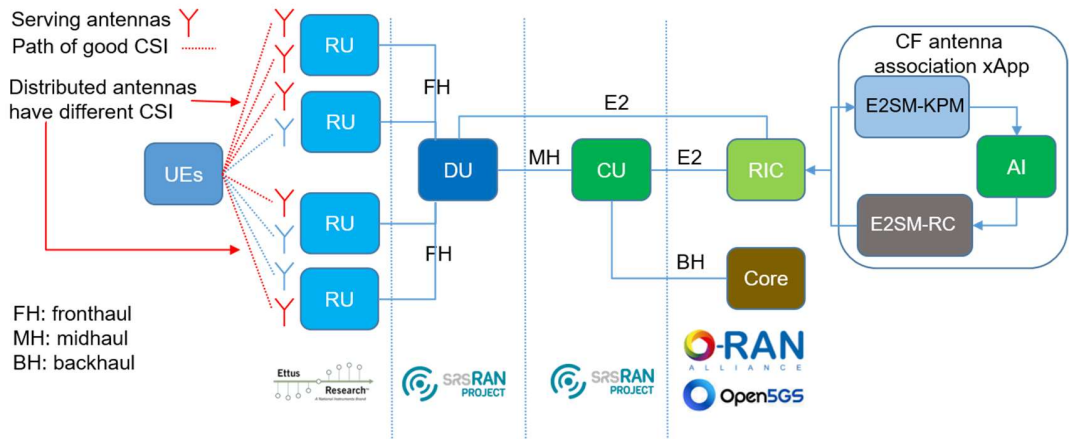

Fig. 1: O-RAN based cell-free testbed architecture

## II. SYSTEM MODEL

The testbed is built purely based on open source components (the codebase can be provided upon request). The testbed is designed to validate the CF architecture in an UL scenario which is relatively less restricted (for RAN modifications) compared with DL when serving COTS UEs (which normally do not support more than 4 antennas). Fig. 1 shows the architecture of the testbed.

**CU and DU**: these are O-RAN compatible and built based on the srsRAN release 24.04[1]. We have modified the DU significantly to implement MU-MIMO and customised E2 functions to support the CF architecture. Detailed modifications to the MAC scheduler, upper PHY and the E2 agent will be described in Section III. The modified DU supports a scalable number of antennas for UL and the maximum number of DL antennas remains original (4). The platform for hosting all software components is a Dell PowerEdge R7525 server (64 cores).

**RIC**: the testbed includes a dockerised near real-time (Near-RT) RIC from the O-RAN software community (OSC)[2]. This is a lightweight version of the original OSC Near-RT RIC Release I provided by the srsRAN team. FlexRIC[3] also partially operates with the testbed, however, not all RIC report styles are supported with the E2 service model (E2SM)-key performance measurement (KPM) which is a typical interoperability issue across O-RAN vendors.

**Intelligent CF antenna association xApp**: this xApp is developed within the OSC RIC. The xApp uses E2SM-KPM to collect standardised and customised KPMs and E2SM - RAN control (E2SM-RC) to control the group of antennas which contribute to the UL. We have pretrained a deep Q network (DQN) model which determines the antenna association based on the KPMs. More details of this xApp will be provided in Section III.

**RU**: we use the National Instruments universal software radio peripheral (USRP) X310 as split 8 RUs. Each RU is equipped with 2T2R antennas. All USRPs are synchronised with a common 10 MHz and 1 PPS signal distributed by an Octoclock. This synchronisation approach limits the deployment distance between RUs (due to attenuation in coaxial cables) and more practical alternatives are N3xx/N4xx radios which can be synchronised via the fibre fronthaul using PTP. The testbed includes 4 RUs which form a 4T8R CF-MIMO network. The number of UL antennas can be further increased if more RUs are available.

---
[1]https://github.com/srsran/srsRAN_Project
[2]https://github.com/srsran/oran-sc-ric
[3]https://gitlab.eurecom.fr/mosaic5g/flexric

**Core network**: we use Open5GS which is a 5G SA core aligned with 3GPP Release 17.

**UE**: unlike many other CF-MIMO testbeds, our testbed supports COTS 5G handsets. The tested handsets include: Xiaomi 12, Xiaomi 11 Lite 5G NE, Xiaomi Redmi Note 10 5G, Google Pixel 6 and OnePlus Nord 5G. The test SIM cards are from Sysmocom.

## III. IMPLEMENTATION DETAILS

The original srsRAN DU supports a single RU per DU (cell) with up to 4T4R antennas. We have modified the DU to allow multiple RUs per DU with up to 4TxR antennas (x indicates scalable). With the 4 RUs in our testbed, all RF chains are in use for the first 2 RUs and only RX chains are in use for the other 2 RUs (for UL). 5G includes 3 physical UL channels, physical random access channel (PRACH), physical uplink control channel (PUCCH) and physical uplink shared channel (PUSCH). Among these the PRACH does not support MIMO, the traffic on PUCCH is low and PUCCH packets are often multiplexed onto PUSCH, so PUSCH becomes our main entry for modifications. The original srsRAN DU supports only SU-MIMO. Upon a UE SR the MAC scheduler arranges an UL grant and ensures no collisions with previously scheduled UL grants. The UL grant is then packed into a packet data unit (PDU) and sent to the PHY via the functional application platform interface. After receiving the IQ of each slot (14 OFDM symbols), the PHY decodes the PUSCH data of each UE sequentially using the information carried by the corresponding PDU such as the UE radio network temporary identifier (RNTI), MCS, allocated PRBs, DMRS type, DMRS symbol masks and port, and slot index. All the above procedures need to be modified for MU-MIMO support, and the detailed modifications are described below.

*A. MAC Scheduler for MU-MIMO Support*

Fig. 2 shows the procedures of the original and modified MAC scheduler. The original scheduler supports only SU-MIMO and ensures no PRB collisions across multiple UEs and the modified scheduler allows a maximum of 2 UEs to share the same PRBs for MU-MIMO. The scheduler allocates PRBs for UEs using round robin on a slot basis and the modifications prioritise MU-MIMO for better spectral efficiency. At every slot the modified scheduler checks all UEs for opportunities of MU-MIMO (regardless new or retransmissions) to maximise the total throughput. The details of the modified scheduler are illustrated below.

The modified MAC scheduler is capable of scheduling multiple UL grants (from multiple UEs) on to the same PRBs while coexisting with SU-MIMO. The original MAC scheduler maintains a bitmap (*bm_su*) where each bit represents whether each resource element (RE) is occupied by a UE. We have modified *bm_su* so that bit 1 indicates the RE is occupied by at least 1 UE. We have added another bitmap (*bm_mu*) to represent whether an RE is occupied by only 1 UE. For *bm_mu*, bit 0 indicates that the RE is occupied by 1 UE and bit 1 indicates that the RE is either not occupied or occupied by 2 UEs. We have also added a vector which stores the UE(s) RNTI(s) allocated to each RE. This vector is used to build the PDU for the PHY.

With every new or pending SR, the scheduler first estimates the required PRBs and MCS based on the UE's channel state information (CSI) and data. The two diamond blocks of low PRB ($\leq 1$ and $\leq 2$) and MCS ($< 6$) exist because the scheduler has issues scheduling UL grants with only 1 PRB while the MCS is low. The solution of srsRAN is to schedule 2 PRBs when such a condition is met and we exclude scheduling such UL grants with MU-MIMO. Otherwise, the scheduler checks *bm_mu* for opportunities of MU-MIMO. If all REs of *bm_mu* are marked as 1 (unsuitable for MU-MIMO), the scheduler checks *bm_su* to find a range of consecutive unoccupied REs that match the request or the longest range of REs possible if there are not enough REs available. Upon successful scheduling, the selected REs in both *bm_su* and *bm_mu* are marked as 1, indicating that these REs can be scheduled to another UE for MU-MIMO.

If the scheduler sees opportunities for MU-MIMO it checks if the conditions in the bottom diamond block is met. If yes, the scheduler skips MU-MIMO scheduling and follows the same steps of using *bm_su* to schedule the UL grant. Otherwise, the scheduler checks *bm_mu* to find a range of consecutive REs available for MU-MIMO that match the request or the longest range of REs possible if there are not enough REs available. Upon successful scheduling, the selected REs in *bm_mu* are marked as 1. These REs should be equal to or a subset of the REs of the first UE allocated to this RE range earlier. The scheduler then finds the REs that are allocated to the first UE and mark them as 1 in *bm_mu*. This avoids the same set of REs being allocated to more than 2 UEs which significantly increases the complexity of PHY.

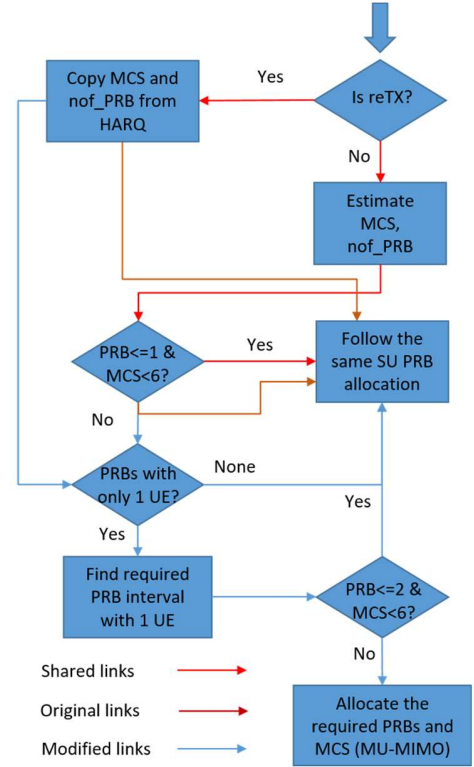

Fig. 2: Original and modified MAC scheduler logic

For retransmissions, the scheduler uses the same MCS and number of PRBs of the previous transmission and checks opportunities for MU-MIMO. The rest of the steps are the same as above. After an UL grant is scheduled, the MAC layer builds a PDU which carries the information the PHY needs to decode the PUSCH data. We have modified the PDU format with two additional elements: *mu_flag* which is a Boolean variable indicating whether this PDU is intended for MU-MIMO, and *rnti_mu* which stores the RNTI of the other UE when *mu_flag* is *True* (the UE scheduled on to unoccupied REs has *False mu_flag*). The MAC layer also builds a downlink control information (DCI) for the UE which carries a similar set of information. Within the DCI the element *antenna_ports* configures the DMRS ports for the UE according to table 7.3.1.1.2-8 in [13]. The original MAC layer configures *antenna_ports*=2 (DMRS port 0) for all UEs and our modification configures a different *antenna_ports* when *mu_flag* of the PDU is *True*. This allows the UEs to use orthogonal DMRS to avoid pilot contamination.

*B. Upper PHY for MU-MIMO Support*

At each slot point, the PHY receives a pool of PDUs (from MAC) corresponding to the UEs scheduled within this slot. The original PHY processes the PDUs sequentially and our modified PHY is capable of processing 2 PDUs (scheduled with MU-MIMO) simultaneously. The PHY first tries to find a PDU with *mu_flag* set to *True*. If such a PDU is not found, the PHY processes all PDUs sequentially. If such a PDU exists, the PHY searches the PDU pool again to identify the other PDU which has its UE RNTI matching *rnti_mu* of the first PDU. These two PDUs are processed together since the channel estimation results of both UEs are needed for the equaliser to separate their IQ. The PHY only processes PDUs with *mu_flag* set to *False* when all PDUs with *mu_flag* set to *True* are processed. We have made significant changes to the PHY's code structure for processing 2 PDUs in parallel but most underlying PHY functions remain unchanged such as carrier and sampling frequency synchronisation, channel estimation, CSI reporting, demapping, descrambling and low-density parity-check decoding. The testbed supports UL MU-MIMO only therefore reciprocity calibration does not arise. The original PHY utilises a maximum-ratio combining (MRC) equaliser, which only supports SU-MIMO. We have integrated a ZF equaliser that leverages the CSI from both UEs to equalise two IQ streams [14].

*C. E2 Agent for CF-MIMO Architecture*

The original srsRAN has a built-in E2 agent which handles the E2 access protocol (E2AP), E2SM-KPM and E2SM-RC messages between the RAN and the RIC. We have made several modifications to E2SM-KPM and E2SM-RC to implement customised RAN functions for the CF-MIMO architecture. To allow the antenna association xApp to select the most suitable antenna group to serve a UE, we have exposed several non-O-RAN standardised KPMs to the E2 agent. These KPMs are antenna specific CSI metrics including the SNR, reference signal received power (RSRP), noise variance and energy per resource element (EPRE). These can be subscribed by any xApp (that supports the standard E2SM-KPM protocol) using metric names *PUSCH.PORT.SNR*, *PUSCH.PORT.RSRP*, *PUSCH.PORT.NVAR* and *PUSCH.PORT.EPRE*. We have also exposed the control of the group of antennas that contribute to the equalisation of the UE IQ. We have defined a customised RAN parameter ID that any xApp (which supports the standard E2SM-RC protocol) can use to configure on/off of each antenna for a specific UE. The DU maintains the configurations for each UE and these configurations are used by the MAC layer when it builds the PHY PDU. We have added a new element *cf_port_selection* in the UE configurations which is an array of binary numbers that represent whether each antenna should contribute to the equalisation. This element is updated every time after the E2 agent receives a RAN Control message from the RIC with the corresponding RAN parameter ID included.

*D. Intelligent Antenna Association xApp*

This example xApp can subscribe to 6 O-RAN standardised KPMs and 4 non-standardised KPMs mentioned in subsection D. All 5 RIC report styles are supported but we prefer style 4 which obtains the KPMs and UE IDs from all UEs. The xApp uses the KPMs as inputs to an embedded DQN which determines the activation or deactivation of a certain antenna for a UE. This actuation is sent to the RAN via E2SM-RC and ultimately changes the UE configuration *cf_port_selection*. The embedded DQN is a lightweight model with 3 hidden convolutional layers, 8×4 input space dimension and 16 action space dimension. The input space consists of the current status of each antenna (aligned with *cf_port_selection*), *PUSCH.PORT.SNR*, *PUSCH.PORT.RSRP* and *PUSCH.PORT.EPRE*. The actions are activating or deactivating a certain antenna. The inference time is generally within the 0.7-2 ms range and one action per UE is needed for every round of KPMs (every second). Once a set of these KPMs are received for a UE, the DQN generates an action and the xApp encodes one RAN Control message. The DQN generates actions individually for each UE and an antenna is only deactivated in the ZF equaliser when both UEs decide to exclude that antenna. Activating more antennas can potentially increase the SNR after equalisation but also increase the processing time (roughly 50 ns more per additional antenna per PDU with our testbed) and the reward is computed based on this trade-off. Processing time is critical for a split 8 system (which has lower and upper PHY on the CPU) and may cause real time issues if not managed properly. We have implemented a simple simulator to pretrained this DQN since training with a live system would take a significant amount of time.

IV. TESTBED DEMONSTRATION

This testbed was implemented as part of the YO-RAN[4] and REACH[5] projects. To allow the readers better understanding the features of the testbed, we have prepared a demonstration video[6] with live commentary showcasing the operations. This section describes the configurations of the experiments in detail, explains the showcased functionalities and presents performance measurements.

*A. Experiment Configurations*

We have conducted the experiments inside the Institute for Safe Autonomy at the University of York. The lab is a 100 m$^2$ indoor lab and the 4 RUs are mounted on the roof rails (Fig. 3). The testbed was configured to operate with a 20 MHz bandwidth on the n78 TDD band (the testbed supports up to 100 MHz bandwidth). The maximum DL MCS was configured as 256 QAM and the maximum UL MCS was configured as 64 QAM (256 QAM was not supported by many COTS UEs). The TDD pattern was configured with 6 slots assigned for DL and 3 slots assigned for UL. The 2 test UEs used in the experiments were a Xiaomi 12 and a OnePlus Nord 5G. The testbed configurations are summarised below in Table III.

*B. Demonstration Walkthrough*

The video demonstration starts with the deployment of the components including the core, the RIC and the CU/DU. A UE was attached to the network to showcase the throughput performance. The RAN metrics showed that the DL throughput reached a maximum of 127 Mbps (tested with iperf3) with a 17 MCS (maximum 28) and 4 layers of data, and the UL throughput reached a maximum of 19 Mbps with a 28 MCS and 1 layer of data (COTS UE supports 1 TX antenna). The combination of higher maximum MCS (256 vs 64 QAM), larger number of slots (6 vs 3) and more layers of data (4 vs 1) of DL has resulted in a much higher DL throughput compared with UL. This maximum of 19 Mbps UL throughput was also the maximum cell throughput for the SU-MIMO case since all the PRBs and UL slots were assigned to this UE. A second UE was then attached to the network to showcase the modified MU-MIMO for UL. The UL iperf tests showed that both UEs had a 16 Mbps throughput which was not achievable with SU-MIMO of the original srsRAN.

---

[4]https://yo-ran.org/
[5]https://www.reach-oran.org/
[6]https://tinyurl.com/52b6pt7y

TABLE II: Testbed Configurations

| Configuration Item | Value |
|---|---|
| Number of DL antennas | 4 |
| Number of UL antennas | 8 |
| Centre frequency | 3600 MHz |
| Channel bandwidth | 20 MHz |
| Subcarrier spacing | 30 kHz |
| Maximum DL modulation | 256 QAM |
| Maximum UL modulation | 64 QAM |
| TDD pattern periodicity | 10 slots |
| Number of UL/DL slots | 3/6 slots |
| RU output power | 15 dBm |
| E2AP version | R003-v03.00 |
| E2SM-KPM version | R003-v03.00 |
| E2SM-RC version | R003-v03.00 |

The demonstration then showcased the intelligent antenna association xApp. The xApp subscribed to 2 O-RAN standardised KPMs (*DRB.UEThpUL* and *DRB.UEThpDL*) and 4 aforementioned non-standardised KPMs. At the beginning all antennas had good CSI and the xApp decided that all antennas should contribute. An anomaly was later manually created to simulate the scenario that the UE was far away from one of the antennas (by disconnecting that antenna). The changes in antenna CSI were observed from the KPMs and the xApp decided to exclude that antenna from serving the UEs. More anomalies were created to showcase that the pretrained DQN was reacting to the environment changes correctly. When the disconnected antennas were reconnected, the xApp decided that these antennas should serve the UEs again.

### C. Performance Measurements

Table III shows the performance measurements when the xApp selected different numbers of antennas (measurements are averaged across a duration of 30 seconds). The measurements include the energy consumption (logged with Power-TOP) of the processor (which hosts CU/DU/core/RIC), total UE throughput and the processing time for 2 MU-MIMO PUSCH PDUs. 2 UEs with iperf were used to generate data for MU-MIMO. Similar to the demo video, we have unplugged different numbers of antennas to allow the xApp to deselect them via E2. With each unnecessary antenna removed from the equaliser, the processor energy consumption was reduced by 1 W (or less) and the PUSCH processing time was reduced by 0.2 us (or less). With each useful antenna included in the equaliser, the total throughput gained 0.4 to 2.6 Mbps. The results show that the xApp can adapt to the changing configuration and improve all three measurements.

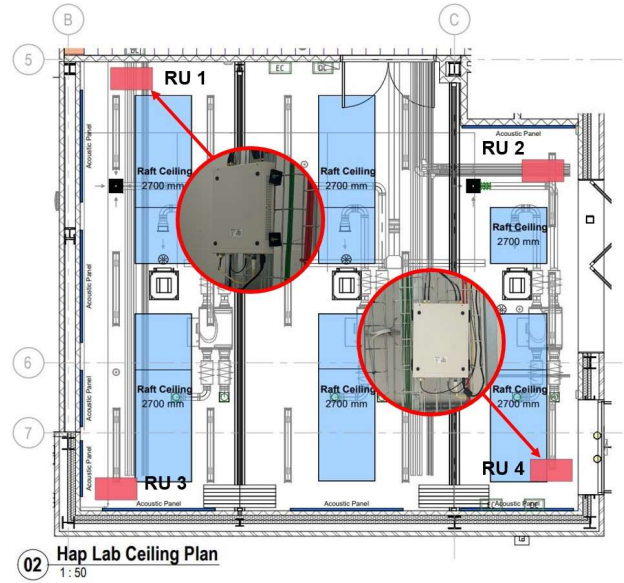

Fig. 3: Testbed deployment

TABLE III: Testbed Performance Measurements

| Antennas selected | Processor power consumption (W) | Total UE throughput (Mbps) | PUSCH processing time (us) |
|---|---|---|---|
| 8 | 58.9 | 38.2 | 4.1 |
| 7 | 58.2 | 37.8 | 3.9 |
| 6 | 57.1 | 36.2 | 3.8 |
| 5 | 56.8 | 35.0 | 3.7 |
| 4 | 55.8 | 33.6 | 3.7 |
| 3 | 55.6 | 32.3 | 3.6 |
| 2 | 54.9 | 29.7 | 3.6 |

## V. CHALLENGES AND FUTURE PATHWAYS

Many challenges remain for implementing the proposed O-RAN based CF network. In this section, we discuss the challenges and future pathways.

1. Network scalability: multiple challenges exist for large-scale CF network deployment including **UE mobility**: although the CF architecture reduces the cell boundaries, in large networks the mobile UEs will eventually cross boundaries of multiple DUs due to the physical limitation of the fronthaul distance and handover is inevitable. Active handover can be implemented to reduce the disruption of service of the UEs. **Dense UEs**: A CF network should support a scalable number of UEs for MU-MIMO according to the number of distributed antennas available, however, the complexity of MAC and PHY escalates rapidly with the increasing number of UEs, particularly if the scheduler allows the allocation of UEs requesting different numbers of REs into the same RE range. In this case the PHY needs to process different combinations of UEs sharing various ranges of REs to obtain the transport blocks for a slot, which significantly increases the complexity.

2. Limited number of orthogonal DMRS: a deterministic aspect for implementing the above bullet point is the number of orthogonal DMRS available. The current srsRAN release supports only type A which includes 8 orthogonal DMRS. When mapping the DMRS ports to antenna ports in the DCI according to the table 7.3.1.1.2-8 in [13], srsRAN only supports number of DMRS CDM group(s) without data set to 2, which further reduces the number of orthogonal DMRS to 4. We have tested all 8 type A DMRS with COTS UEs and only antenna ports 2, 4 and 5 have correct channel estimation. In practice up to 3 UEs can be supported with MU-MIMO unless the DMRS support is updated.

3. Processing time for MU-MIMO: the original srsRAN uses the least time consuming MRC equaliser. Our ZF equaliser has doubled the processing time per PDU due to various matrix operations (using the Eigen library) and real time issues may appear with larger bandwidth. The equaliser can be further optimised to improve the stability of the RAN.

4. More RAN Control via E2: more configurable parameters can be exposed to support optimisations specific to CF such as **UL UE power control**: the PUSCH power is determined by many factors such as the configurable *P0*, the number of PRBs, the MCS and the path loss. An xApp can be designed to allocate the appropriate UL power according to the UE KPMs and MIMO modes. **DU/CPU location optimisation**: it may be beneficial to include as many RUs as possible at a DU/CPU, however, the distance of the fronthaul cannot be extended indefinitely due to the signal latency in fibre. The AI within the RIC can potentially determine the location to deploy the DU/CPU which provides the optimal network performance. Interoperability should be considered when expanding RAN Control functions to allow RIC/xApp from multiple vendors to operate directly. The proposed testbed has exposed customised KPMs and RAN parameters via O-RAN standardised methods and such principle should persist when implementing new functions.

## VI. Conclusions

This paper has presented the design and implementation of an O-RAN based CF-MIMO network and proposed a novel hybrid network architecture with the use of AI. The presented testbed included a full 5G stack and a video has demonstrated the network serving COTS 5G UEs. Key contributions include the implementation of MU-MIMO support, an intelligent antenna association xApp using AI-driven optimizations, and critical modifications to the MAC and PHY layers of the open source srsRAN. We have also discussed the challenges for improving the capability of the testbed, as well as unique optimisation opportunities in the RIC.

## Acknowledgements

This work was supported by project Yorkshire Open Ran (YO-RAN) and project RIC Enabled (CF-)mMIMO For HDD (REACH) funded by the Department for Science, Innovation and Technology (DSIT) of the UK government.